\documentclass[12pt]{article}
\usepackage[utf8]{inputenc}
\usepackage{graphicx}
\usepackage{amsmath}
\usepackage{hyperref}
\usepackage{geometry}
\usepackage{cite}
\usepackage{multirow}
\usepackage{amsmath, amsfonts, amssymb} 
\usepackage{graphicx} 
\usepackage{multirow} 

\title{Enhancing Brain Tumor Segmentation Using Channel Attention and Transfer learning }
\author{
    Majid Behzadpour\thanks{Department of Electrical and Computer Engineering, University of Tehran, Tehran, Iran. Email: majid.behzadpour11@gmail.com} \and Ebrahim Azizi\thanks{Department of Electrical and Computer Engineering, Texas Tech University, Lubbock, TX, USA.} 
     \and 
        Kai Wu\thanks{Department of Electrical and Computer Engineering, Texas Tech University, Lubbock, TX, USA. Email: kai.wu@ttu.edu} 
     \and 
    Bengie L. Ortiz\thanks{Department of Pediatrics, Hematology and Oncology Division, Michigan Medic
    ine, University of Michigan Health System, Ann Arbor, MI, USA.}
}

\date{}

\begin{document}

\maketitle

\begin{abstract}
Accurate and efficient segmentation of brain tumors is critical for diagnosis, treatment planning, and monitoring in clinical practice. In this study, we present an enhanced ResUNet architecture for automatic brain tumor segmentation, integrating an EfficientNetB0 encoder, a channel attention mechanism, and an Atrous Spatial Pyramid Pooling (ASPP) module. The EfficientNetB0 encoder leverages pre-trained features to improve feature extraction efficiency, while the channel attention mechanism enhances the model's focus on tumor-relevant features. ASPP enables multiscale contextual learning, crucial for handling tumors of varying sizes and shapes. The proposed model was evaluated on two benchmark datasets: TCGA LGG and BraTS 2020. Experimental results demonstrate that our method consistently outperforms the baseline ResUNet and its EfficientNet variant, achieving Dice coefficients of 0.903 and 0.851 and HD95 scores of 9.43 and 3.54 for whole tumor and tumor core regions on the BraTS 2020 dataset, respectively. compared with state-of-the-art methods, our approach shows competitive performance, particularly in whole tumor and tumor core segmentation. These results indicate that combining a powerful encoder with attention mechanisms and ASPP can significantly enhance brain tumor segmentation performance. The proposed approach holds promise for further optimization and application in other medical image segmentation tasks.
\end{abstract}


\section{Introduction}
Brain tumors arise as a result of the uncontrollable proliferation of abnormal cells. Although their causes are not well known, some of the risk factors include family history, metastases, and exposure to ionizing radiation \cite{icsin2016review}. A very common type of brain tumor is glioma. The world health organize has divided glioma into two categories: low-grade (Grade I and Grade II) gliomas and high-grade (Grade III and Grade IV) gliomas. Low-grade gliomas (LGG) tend to be less aggressive, with longer survival rates, while high-grade gliomas (HGG) exhibit rapid progression and require urgent medical intervention, with a median survival rate of fewer than two years \cite{menze2014multimodal}.

Magnetic Resonance Imaging(MRI) is generally considered the best imaging technique in brain tumor diagnosis due to the non-invasiveness and high resolution it affords, aside from being able to provide several types of structural, functional, and diffusion MRI in detail to examine anatomy and tumor characteristics of the brain \cite{bauer2013survey}. For diagnostic, treatment, or monitoring purposes, the correct and prompt segmentation of tumors from brain MRI assumes paramount importance. While manual segmentation performed by radiologists remains the clinical standard, it is a labor-intensive, time-consuming process that is prone to intra- and inter observer variability \cite{icsin2016review}, \cite{yusoff2023accuracy}. Hence, there is an increasing demand for automated brain tumor segmentation techniques. The challenge in segmenting brain tumors lies in the difference of shape, size, and location each tumor assumes from one patient to another. Most high-grade gliomas invade the surrounding normal-appearing tissues, hence their borders cannot be well delineated. Additionally, MRI images obtained from various modalities such as flair, T1, T1ce, and T2 differ in their degree of contrast, making segmentation even more challenging \cite{mandal2012structural}, \cite{pellico2019nanoparticle}. Traditional machine learning techniques relied on manual feature extraction and labeled datasets for training classifiers; however, these usually faced ambiguous boundaries between malignant and benign tissues and needed substantial expert input at the initial annotation stage \cite{mandal2012structural}.

Recent advancements in deep learning have revolutionized brain tumor segmentation. Convolutional neural networks (CNNs) and fully convolutional networks (FCNs) have demonstrated remarkable success in medical image analysis due to their ability to learn complex hierarchical features from large datasets. Notably, the introduction of the Brain Tumor Segmentation Challenge (BraTS) has accelerated the development of robust, standardized models for brain tumor segmentation by providing high-quality annotated datasets \cite{pellico2019nanoparticle}. U-Net and its variants have become the most widely adapted deep models in brain tumor segmentation, often achieving state-of-the-art performance compared to other deep learning-based models \cite{lefkovits2022u}. However, such networks usually have a few limitations: they are not good particularly effective at modeling complex contextual features or processing multi-scale information efficiently.

In this work, we propose an improved ResU-Net architecture for brain tumor segmentation using multimodal MRI data. Our model is based on some previous works in ResU-Net-based architectures and embodies a number of key improvements to enhance the accuracy of segmentations:

\hangindent=1cm
•   Channel Attention Mechanisms: The traditional U-Net model regards all channels with equal importance, so we will use a channel attention mechanism. This approach enables the network to selectively emphasize the most informative channels of each MRI modality for enhancing the focus on tumor regions. 

\hangindent=1cm
•   Atrous Spatial Pyramid Pooling (ASPP): An ASPP module is integrated at the bottleneck to capture multi-scale contextual information, which is quite essential for the segmentation of tumors accurately with varied sizes and shapes. ASPP lets the model embed both local and global contextual clues \cite{srinanda2024cascrnet}.

We further enhance our approach by leveraging EfficientNet as the backbone encoder, which provides a powerful feature extraction capability while maintaining computational efficiency \cite{tan2019efficientnet}. Its compound scaling strategy ensures a balanced trade-off between model depth, width, and resolution, allowing for better representation of complex brain tumor features with reduced computational costs.

We evaluate our model on two benchmark datasets: the TCGA LGG dataset \cite{mazurowski2017radiogenomics} containing 3929 MRI scans of 110 patients and BraTS2020 \cite{baid2021rsna}. The TCGA dataset would better represent real-world variations, while BraTS2020 includes standardized ground truth annotations and is, therefore, more suitable for benchmarking segmentation models. Our approach targets state-of-the-art performance by addressing the key limitations of existing models through advanced architectural modifications. The code for our implementation is publicly available at \href{https:// https://github.com/majid9418/EResU-Net}{GitHub}.

\section{Related Works}
The field of deep learning models for brain tumor segmentation has undergone extensive exploration. Deep neural networks have recently become the favored choice, dominating most works on BraTS, particularly since larger datasets have been beneficial to them. Kamnitsas et al. \cite{kamnitsas2018ensembles, kamnitsas2017efficient, kamnitsas2015multi} proposed an ensemble approach, combining predictions from multiple 3D convolutional networks, including DeepMedic, FCN, and U-Net. Similarly, Myronenko et al. \cite{myronenko20193d} investigated a cascaded U-Net model, where a coarse segmentation obtained from the first stage was further refined by a second-stage U-Net. Isensee et al. applied nnU-Net, a self-configuring method that adapts U-Net to particular datasets, achieving state-of-the-art performance in the BraTS challenge \cite{isensee2021nnu}. Further variants of nnU-Net, including the use of group normalization instead of batch normalization and the addition of axial attention, have presented higher segmentation accuracy \cite{luu2021extending}.

The attention mechanism has played a pivotal role in improving brain tumor segmentation. Noori et al. \cite{noori2019attention} applied channel attention immediately after the concatenation of low-level and high-level features in a 2D encoder-decoder architecture, emphasizing the importance of weighing different features over naive concatenation. Zhang et al. \cite{zhang2020attention} introduced attention gates at the skip connections, further improving segmentation accuracy. Additionally, Cao et al. \cite{cao2023mbanet} proposed a U-Net-like model using 3D shuffle attention in the encoder for better feature extraction and skip connections. These studies highlight how the inclusion of attention mechanisms enhances segmentation quality.

U-Net is one of the foundational architectures for segmenting medical images. The model proposed by Ronneberger et al. \cite{ronneberger2015u} includes a contracting path for feature extraction and an expansive path for precise localization through upsampling. The concatenation of encoder and decoder features through skip connections enhances segmentation accuracy by preserving spatial details. Furthermore, hybrid models, such as U-Net combined with other architectures like ResNet, have achieved excellent results in brain tumor segmentation tasks, with Dice scores higher than 0.90 on benchmark datasets \cite{ronneberger2015u, he2016deep, saha2021brain}. Recent works also feature a genetic algorithm-based CNN that automatically optimizes the network architecture, achieving accuracy up to 0.94 on the TCGA dataset \cite{anaraki2019magnetic}.

Further developments in brain tumor segmentation involve modifications of the standard U-Net model for better performance. Baid et al. \cite{baid2019deep, iqbal2022brain}presented a 3D patch-based U-Net for glioma segmentation and survival prediction. Their approach, using a Deep Learning Radiomics Algorithm for Gliomas (DRAG), achieved significant results, with Dice coefficient values of 0.9795 for high-grade gliomas and 0.9950 for low-grade gliomas. Another noteworthy approach is the Spatial Pyramid Pooling U-Net (SPP-U-Net), which integrates spatial pyramid pooling with attention blocks to improve multi-scale feature representation \cite{vijay2023mri}. Additionally, researchers like Sunita Roy et al. \cite{roy2023brain} have introduced advanced architectures, such as S-Net and SA-Net, that utilize attention-based mechanisms to enhance segmentation results on MRI scans.

Recent advancements also explore new architectures beyond traditional CNNs. For instance, Ruba et al. \cite{ruba2023brain} proposed JGate-AttResU-Net, which incorporates an attention gate and residual blocks to reliably highlight variably sized tumor regions. Yanjun Peng and Jindong Sun \cite{peng2023multimodal} proposed an efficient automatic weighted dilated convolutional network (AD-Net) architecture, based on weighted dilated convolutions and deep supervision, to enhance segmentation efficiency. More recently, transformer-based models like TransBTS \cite{wang2023learning} have shown that attention-driven models are capable of learning robust features for brain tumor segmentation. These works highlight the trend of using more complex neural network elements, including transformers and multi-scale feature fusion, to achieve better segmentation results.

\section{Methodology}
In this work, we propose an improved version of the ResU-Net model as the backbone in our approach to brain tumor segmentation. First, we propose the structure of ResU-Net, enhanced by the EfficientNetB0 encoder to leverage pre-trained features and improve efficiency. Then we introduce the addition of Atrous Spatial Pyramid Pooling (ASPP) in the bottleneck to model multi-scale contextual information. The channel attention mechanisms followed are then elaborated, showing how they serve to help in feature selection by emphasizing the relevant regions. Next, the integration of residual blocks at every place in the network for capturing better feature flow and refinement is introduced. Finally, an outline of the training strategy adopted to optimize the network for accurate segmentation is described.

\subsection{Baseline Model: ResU-Net with EfficientNetB0 Encoder}
Our baseline model relies on the ResU-Net architecture \cite{diakogiannis2020resunet} incorporating an EfficientNetB0 encoder \cite{tan2019efficientnet}. The network has an encoder-decoder structure with skip connections between corresponding layers. The multi-modal MRI inputs are processed by a pre-trained EfficientNetB0 backbone initialized with ImageNet weights \cite{krizhevsky2017imagenet} that can be further fine-tuned during training.

The encoder pathway utilizes four key activation layers from EfficientNetB0 (block1a through block4a) to extract hierarchical features. While the pathway in the decoder comprises a total of four upsampling levels, reducing feature channels from 256 to 32 through transposed convolutions with stride 2. At each level, decoder receives skip connections from the corresponding encoder activation layer, followed by a residual block for  feature refinement \cite{he2016deep}. 

This final output layer consists of a 2 × 2 transposed convolution that upsamples to the input resolution, followed by a 1 × 1 convolution with sigmoid activation to generate three output channels corresponding to whole tumor, tumor core, and enhancing tumor regions. 

\subsection{Channel Attention Mechanism}
Our network incorporates channel attention mechanisms in the decoder pathway to adaptively re-calibrate channel-wise feature responses. Unlike spatial attention that focuses on specific regions, our channel attention mechanism emphasizes informative feature channels while suppressing less important ones. The attention module is applied after each skip connection concatenation in the decoder pathway to effectively calibrate the merged features from different scales. 

Given an input feature map \( F \in \mathbb{R}^{H \times W \times C} \), the channel attention module first generates two different channel descriptors through global average pooling \( F_{\text{avg}} \) and global max pooling \( F_{\text{max}} \) operations:

\[
F_{\text{avg}} = \frac{1}{{H \times W}} \sum_{i=1}^{H} \sum_{j=1}^{W} F(i,j) \tag{1}
\]

\[
F_{\text{max}} = \max_{(i,j)} F(i,j) \tag{2}
\]

where \( F(i,j) \) represents the feature value at spatial position \( (i,j) \). The channel attention map \( M_c \) is then computed by:

\[
M_c = \sigma \left( F_{\text{avg}} + F_{\text{max}} \right) \tag{3}
\]

where \( \sigma \) denotes the sigmoid activation function. The final output \( F' \) is obtained by rescaling the input feature map with the attention weights:

\[
F' = M_c \odot F \tag{4}
\]

where \( \odot \) denotes channel-wise multiplication. The attention mechanism is systematically integrated at four levels in the decoder pathway, processing feature maps with varying channel depths (256, 128, 64, and 32 channels). This hierarchical application of channel attention helps in refining features at multiple scales, particularly after the fusion of skip connections with upsampled features, leading to more discriminative feature representations for accurate tumor segmentation.

\subsection{Atrous spatial pyramid pooling (ASPP)}
In particular, we embed an Atrous Spatial Pyramid Pooling (ASPP) module \cite{chen2017deeplab} at the bottleneck of the network to capture multi-scale contextual information that is critical for segmenting brain tumors of diverse sizes and shapes. This ASPP module performs parallel atrous convolutions at different dilation rates of 6, 12, and 18, along with a 1 × 1 convolution branch and a global pooling branch. Each branch processes the input features using 256 filters, followed by ReLU activation. Features from all the branches are then concatenated and processed through a final 1×1 convolution, hence letting the network capture contextual information effectively at different scales with no loss in spatial resolution.
    
\section{Experiments}
\subsection{Datasets and Augmentation}
We utilized two datasets: BraTS-Glioma dataset [12] provided by Brain Tumor Segmentation Challenge 2023, and the TCGA LGG dataset, containing 3929 MRI scans of 110 patients. The BraTS MRI scans in the BraTS-Glioma comprise native(T1), post-contrast T1-weighted (T1Gd), T2-weighted (T2), and T2 Fluid Attenuated Inversion Recovery (T2-FLAIR). The primary evaluations have been performed for three tumor sub-regions of interest: the "enhancing tumor" with hyper-intensity in T1Gd (ET), the "tumor core" that includes both necrotic regions and ET (TC), and the "whole tumor" that includes TC and also peritumoral invaded tissue (WT). From the available BraTS data, we used 1100 cases for development and 151 cases for testing. From the TCGA LGG dataset, we adopted 28 cases for validation and 28 cases for test set (ratio of 70\% for training, 15\% for validation and 15\% for testing). Models are trained using many augmentation techniques with the goal of increasing robustness and preventing overfitting by using random rotation of ±25 degrees that can capture different types of invariances or orientation variability. It includes using random horizontal/vertical shifts that are within 20\% of the dimensions of an image, random zoom transformations with ±20\% limits, random horizontal flip, and interpolating newly created pixels with one of the following techniques during the above operations: nearest neighbors. These augmentation parameters are selected with care to reflect real variations in medical imaging while preserving anatomical validity in the brain structures.

This study was conducted on Google Colab using Python 3.9 with TensorFlow 2.12 and Keras. The Colab environment provided access to NVIDIA Tesla T4 GPUs, offering sufficient computational power for training deep learning models. The model compilation utilized the 'adam' optimizer with a categorical cross-entropy loss function for multi-class classification, running under CUDA 11.2 to accelerate GPU processing. Additionally, Google Colab’s backend facilitated memory and processing optimizations necessary for handling large histopathological image datasets.
\subsection{Evaluation Metrics}
to evaluate the performance of our brain tumor segmentation models, two common metrics were adopted: The Dice Similarity Coefficient-DSC, and the 95th percentile Hausdorff Distance-HD95, both widely used in medical image analysis. The Dice coefficient refers to the ratio of overlapping between the predicted tumor area and the ground truth annotation, it is defined as:

\[
\text{DSC} = \frac{2 |P \cap G|}{|P| + |G|} \tag{5}
\]

where \( P \) is the predicted tumor region and \( G \) is the region of the ground truth. A higher Dice score indicates a better overlap between the real and predicted masks of tumors.

Complementing the DSC, we also use the HD95 , which is the robust version of the Hausdorff Distance computing the 95th percentile of the maximum surface-to-surface distance between predicted segmentation and ground truth boundary. This metric provides accuracy of the boundary localization, insensitive to extreme outliers by excluding the worst 5\% of boundary mismatches. These two metrics together, Dice coefficient, and HD95 provide a comprehensive measure of volumetric overlap and boundary precision of the segmented tumor regions.

\subsection{Results and Discussion }
In this section, we present a detailed analysis and comparison of the quantitative results obtained from the baseline ResUNet, ResUNet with EfficientNet backbone, and our proposed enhanced method incorporating channel attention and ASPP mechanisms. The evaluation was performed on two distinct datasets: TCGA LGG and BraTS 2020.

\subsubsection{Comparison of the proposed method}
Quantitative results. Tables 1 and 2 show the comparisons of our proposed method with the baseline approaches. The proposed improved model consistently outperforms the different variations of baseline methods for both datasets. On the TCGA LGG dataset, the results depict that our proposed method provides the top segmentation performance with a Dice score of 0.797 and HD95 of 6.16, which is considerably improved from the baseline ResUNet (DSC: 0.787, HD95: 7.21). Its combination with the EfficientNet backbone and data augmentation demonstrated a mediocre boost: DSC 0.781, HD95 - 6.39, whereas our final method further improved the segmentation results by including channel attention and ASPP.

\begin{table}[ht]
\centering
\caption{Compared segmentation results with baselines on TCGA LGG test dataset.}
\begin{tabular}{|l|c|c|c|}
\hline
\textbf{Methods} & \textbf{Aug} & \textbf{DSC} & \textbf{HD95} \\
\hline
ResUnet &  & 0.787 & 7.21 \\
ResUnet & \checkmark & 0.773 & 7.29 \\
ResUnet + EfficientNet &  & 0.775 & 6.47 \\
ResUnet + EfficientNet & \checkmark & 0.781 & 6.39 \\
Our method & \checkmark & 0.797 & 6.16 \\
\hline
\end{tabular}
\end{table}

\begin{table}[ht]
\centering
\caption{Compared segmentation results with baselines on BraTS2020 test dataset.}
\begin{tabular}{|l|c|c|c|c|c|c|c|}
\hline
\textbf{Methods} & \textbf{Aug} & \multicolumn{3}{c|}{\textbf{DSC}} & \multicolumn{3}{c|}{\textbf{HD95}} \\
\hline
 & & \textbf{ET} & \textbf{WT} & \textbf{TC} & \textbf{ET} & \textbf{WT} & \textbf{TC} \\
\hline
ResUnet & & 0.722 & 0.851 & 0.767 & 7.20 & 11.65 & 5.54 \\
ResUnet & \checkmark & 0.725 & 0.863 & 0.770 & 7.18 & 11.13 & 6.21 \\
ResUnet + EfficientNet & & 0.741 & 0.877 & 0.781 & 6.25 & 10.41 & 5.34 \\
ResUnet + EfficientNet & \checkmark & 0.750 & 0.889 & 0.789 & 6.35 & 10.08 & 5.14 \\
Our method & \checkmark & 0.762 & 0.903 & 0.794 & 5.87 & 9.43 & 3.54 \\
\hline
\end{tabular}
\end{table}
On the BraTS 2020 dataset, the proposed approach reports the best performance in the three tumor regions: ET, WT, and TC. The improved model achieves Dice scores of 0.762, 0.903, and 0.851 for ET, WT, and TC respectively, significantly outperforming both a baseline ResUNet and its EfficientNet variant. HD95 also improved significantly, most notably in the TC region, where our approach achieves 3.54 against the baseline value of 5.54. The most difficult part for segmentation, the whole tumor area, demonstrates a much-improved Dice score (0.903) and low HD95 of 9.43 against the baseline of 0.851 and 11.65, respectively.

The impact of data augmentation varies across different model configurations. While the baseline ResUNet shows mixed results with augmentation (slight decrease in TCGA LGG performance but improvement in BraTS 2020), the EfficientNet backbone consistently improved from augmentation across both datasets. This could indicate that the more complex architecture is better at using the augmented data in order to learn robust features.

Figure 1. shows representative segmentation results comparing our proposed method with the baseline approaches. The visual comparison demonstrates that our enhanced model, incorporating channel attention and ASPP modules, produces more precise tumor boundaries that closely align with the ground truth, particularly in areas where the baseline ResUNet struggles to maintain boundary consistency. This visual improvement validates the quantitative results and highlights the effectiveness of our architectural enhancements, especially the synergistic effect of channel attention for feature refinement and ASPP for multi-scale context integration.

The superior performance of our proposed method can be attributed to several factors. The channel attention mechanism helps the model focus on the most relevant features in each modality, while the ASPP module effectively captures multi-scale contextual information. When combined with the EfficientNet backbone and data augmentation, these components work synergistically to achieve more accurate and reliable segmentation results.

\begin{figure}[ht]
    \centering
    \includegraphics[width=0.8\textwidth]{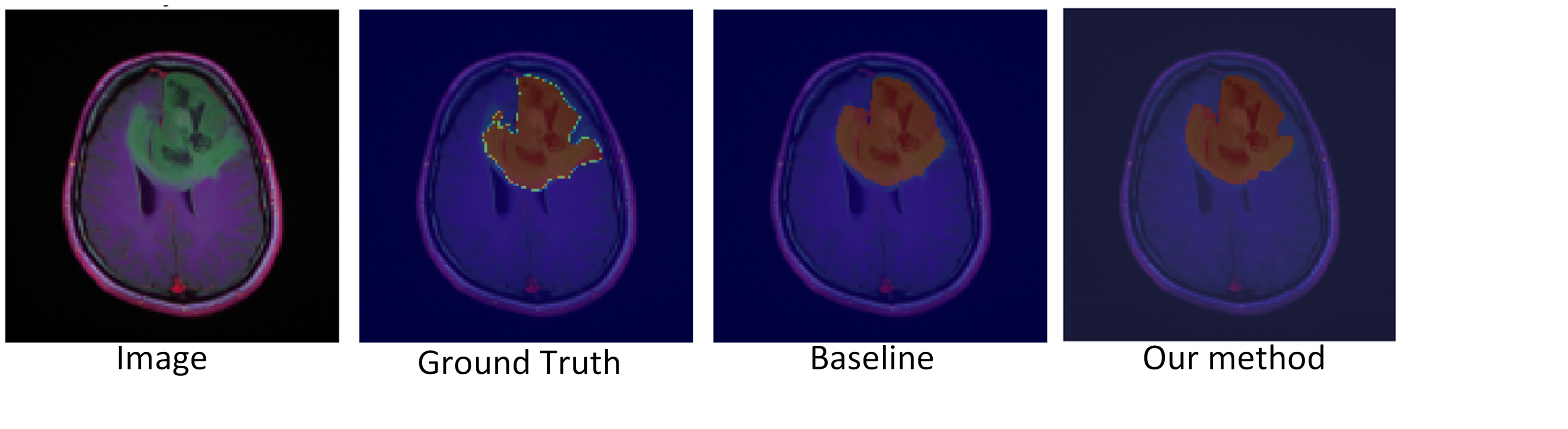} 
    \caption{Segmentation result provided by standard ResU-Net and Our method. This result was obtained from the test set. }
    \label{fig:workflow}
\end{figure}
\subsubsection{Comparison with the state-of-the-art methods}
For a fair comparison, Table 3 presents the performance metrics of our method against several state-of-the-art approaches on BraTS. The comparison includes both recent and established methods: ResU-net, nnU-Net, SCU-Net \cite{zheng2022brain}, ATST \cite{wang2023learning}, and Daza et al. \cite{daza2021cerberus}.

Wang et al. \cite{wang2023learning} (ATST) ranked second during the BraTS challenge by introducing a new method which relied on two interlinked pathways, each processing a pair of modalities. Its method turned out to be quite efficient, particularly on whole tumor segmentation, where its DSC score was 0.9017. Similarly, Daza et al. \cite{daza2021cerberus} proposed a light-weight approach for which showed best performance in various metrics and resulted in the highest DSC values of 0.794, 0.897, and 0.845 for ET, WT, and TC, respectively.

Our method achieves competitive performance: 0.903 for whole tumor segmentation and 0.851 for TC segmentation, achieving a DSC of 0.762 for enhancing tumor comparable to baseline methods such as ResU-net with 0.722 and nnU-Net at 0.720, though inferior compared to recent approaches like ATST and Daza et al. HD95 is rather inconsistent for the different tumor regions, with our method obtaining 5.87, 9.43, and 3.54 for ET, WT, and TC, respectively.

\begin{table}[ht]
\centering
\caption{Segmentation comparison of our method with state-of-the-works. “WT” stands for whole tumorous area, “ET” stands for enhancing tumor, and “TC” stands for non-enhancing components.}
\begin{tabular}{|l|c|c|c|c|c|c|}
\hline
\textbf{Methods} & \multicolumn{3}{c|}{\textbf{DSC}} & \multicolumn{3}{c|}{\textbf{HD95}} \\
\hline
 & \textbf{ET} & \textbf{WT} & \textbf{TC} & \textbf{ET} & \textbf{WT} & \textbf{TC} \\
\hline
ResU-net & 0.722 & 0.851 & 0.767 & 7.20 & 11.65 & 5.54 \\
nnU-Net & 0.720 & 0.810 & 0.770 & 4.2 & 3.5 & 5.5 \\
SCU-Net\cite{zheng2022brain} & 0.713 & 0.791 & 0.772 & 4.1 & 5.1 & 5.4 \\
ATST\cite{wang2023learning} & 0.785 & 0.9017 & 0.837 & 32.25 & 4.39 & 8.34 \\
Daza et al.\cite{daza2021cerberus} & 0.794 & 0.897 & 0.845 & 29.82 & 3.59 & 6.47 \\
Our method & 0.762 & 0.903 & 0.851 & 5.87 & 9.43 & 3.54 \\
\hline
\end{tabular}
\end{table}

\section{Conclusion}
In this study, we proposed an integrated method based on ResUNet for segmenting brain tumors by incorporating an EfficientNetB0 encoder, channel attention mechanism, and Atrous Spatial Pyramid Pooling to reinforce the ability of multi-scale feature extraction and contextual learning. These demonstrated their performances on two benchmark datasets: TCGA LGG and BraTS 2020, and outperformed the baseline ResUNet and its EfficientNet variant. The proposed approach shows the best performances in terms of Dice coefficients and HD95 scores on essential tumor regions, and it reaches a Dice score of 0.903 and HD95 of 9.43 for the whole tumor on BraTS 2020. Compared with the state-of-the-art, our model exhibited competitive performance in whole tumor and tumor core segmentation, validating the effectiveness of combining a powerful encoder with an attention mechanism and ASPP for accurate boundary delineation. These results demonstrate that our approach could perform exceedingly well for clinical practice, and its further optimization and adaptation may be carried out on other segmentation medical tasks.

\bibliography{references.bib}
\bibliographystyle{plain}

\end{document}